\documentclass[fleqn,twoside]{article}
\usepackage{espcrc2}

\usepackage{graphicx}
\usepackage[figuresright]{rotating}

\newcommand{\AmS}{{\protect\the\textfont2
  A\kern-.1667em\lower.5ex\hbox{M}\kern-.125emS}}

\newcommand\mzon   {M$_{\odot}$}

\def\degr{\hbox{$^\circ$}}

\newcommand\Lunit {ergs s$^{-1}$}

\def\la{\mathrel{\mathchoice {\vcenter{\offinterlineskip\halign{\hfil
$\displaystyle##$\hfil\cr<\cr\noalign{\vskip1.5pt}\sim\cr}}}
{\vcenter{\offinterlineskip\halign{\hfil$\textstyle##$\hfil\cr<\cr
\noalign{\vskip1.0pt}\sim\cr}}}
{\vcenter{\offinterlineskip\halign{\hfil$\scriptstyle##$\hfil\cr<\cr
\noalign{\vskip0.5pt}\sim\cr}}}
{\vcenter{\offinterlineskip\halign{\hfil$\scriptscriptstyle##$\hfil
\cr<\cr\noalign{\vskip0.5pt}\sim\cr}}}}}
\def\ga{\mathrel{\mathchoice {\vcenter{\offinterlineskip\halign{\hfil
$\displaystyle##$\hfil\cr>\cr\noalign{\vskip1.5pt}\sim\cr}}}
{\vcenter{\offinterlineskip\halign{\hfil$\textstyle##$\hfil\cr>\cr
\noalign{\vskip1.0pt}\sim\cr}}}
{\vcenter{\offinterlineskip\halign{\hfil$\scriptstyle##$\hfil\cr>\cr
\noalign{\vskip0.5pt}\sim\cr}}}
{\vcenter{\offinterlineskip\halign{\hfil$\scriptscriptstyle##$\hfil
\cr>\cr\noalign{\vskip0.5pt}\sim\cr}}}}}

\title{An observational review of accretion-driven millisecond X-ray pulsars}

\author{Rudy Wijnands\address{School of Physics \& Astronomy, University 
        of St Andrews\\ North Haugh, St Andrews, Fife, KY16 9SS, Scotland, UK}}

\begin{document}

\begin{abstract} I present an observational review of the five
currently known accretion-driven millisecond X-ray pulsars. A
prominent place in this review is given to SAX~J1808.4--3658; it was
the first such system discovered and currently four outbursts have
been observed from this source. This makes SAX J1808.4--3658 the best
studied example of the group.  Its most recent outburst in October
2002 is of particular interest because of the discovery of two
simultaneous kilohertz quasi-periodic oscillations and nearly
coherent oscillations during type-I X-ray bursts. This is the first
time that such phenomena are observed in a system for which the
neutron star spin frequency is exactly known. The other four systems
were  discovered within the last two years and only limited results
have been published. Since new exiting results are to be expected in
the future for all five sources, this review will only represent a
snap-shot of the current observational knowledge of accretion-driven
millisecond X-ray pulsars. A more extended and fully up-to-date
review can be found at http://zon.wins.uva.nl/$\sim$rudy/admxp/.

\vspace{1pc}
\end{abstract}


\maketitle

\section{Introduction}

Pulsars are born as highly-magnetised ($B\sim 10^{12}$ G), rapidly
rotating ($P\sim 10$ ms) neutron stars which spin down on timescales
of 10 to 100 million years due to magnetic dipole radiation. 
However, a number of millisecond ($P<10$ ms) radio pulsars is known
with ages of  billions of years and weak ($B\sim 10^{8-9}$ G) surface
magnetic fields.  Since many of these millisecond pulsars are in
binaries, it has long been suspected (see, e.g., \cite{bvdh1991} for
an extended review) that they were spun up by mass transfer from a
stellar companion in a low-mass X-ray binary (LMXB), but years of
searching for coherent millisecond pulsations in LMXBs failed to
yield a detection \cite[and references therein]{v1994}.  The launch
of the {\itshape Rossi X-ray Timing Explorer} ({\itshape RXTE})
brought the discovery of kilohertz quasi-periodic oscillations (kHz
QPOs; \cite{strohmayer1996,vdk1996}) as well as nearly coherent
oscillations ('burst oscillations') during type-I X-ray bursts in a
number of LMXBs (e.g., \cite{strohmayer1996}), providing
tantalisingly suggestive evidence for weakly magnetic neutron stars
with millisecond spin periods (see \cite{vdk2000} and \cite{sb2003}
for more details about kHz QPOs and burst oscillations in LMXBs).

In 1998 April the first accretion-driven millisecond X-ray pulsar (SAX
J1808.4--3658) was discovered \cite{wvdk1998} proving that indeed
neutron stars in LMXBs can spin very rapidly. This conclusion was
further strengthened by the discovery of four additional systems
during the last two years
\cite{markwardt2002,galloway2002,markwardt2003_1807,markwardt2003_1814}. Here,
I will give a brief summary of our current observational knowledge of
those accretion-driven millisecond X-ray pulsars.

\section{SAX J1808.4--3658}

\subsection{The 1996 September outburst}

In 1996 September, a new X-ray transient and LMXB was discovered with
the Wide Field Cameras (WFCs) aboard the {\it BeppoSAX} satellite and
the source was designated SAX J1808.4--3658 \cite{intzand1998}. Three
type-I X-ray bursts were detected, demonstrating that the compact
object in this system is a neutron star. From those bursts, a
distance estimate of 2.5 kpc was determined
\cite{intzand1998,intzand2001}. The maximum luminosity during this
outburst was $\sim10^{36}$ \Lunit, significantly lower than the peak
outburst luminosity of 'classical' neutron star transients. This low
peak luminosity showed that the source was part of the growing group
of faint neutron-star X-ray transients \cite{heise1999}.  The
outburst continued for about three weeks, after which the source was
thought to have returned to quiescence.  However, recently it was
found \cite{revnivtsev2003} that the source was detected on 1996
October 29 (using slew data obtained with the proportional counter
array [PCA] aboard {\itshape RXTE}) with a luminosity of about a
tenth of the outburst peak luminosity.  This demonstrates that six
weeks after the main outburst the source was still active (possible
only sporadically), which might indicate behaviour for this source at
the end of this outburst very similar to what has been seen during
its 2000 and 2002 outbursts (\S~\ref{section:1808_2000_outburst} and
\ref{section:1808_2002_outburst}).

After it was found that SAX J1808.4--3658 harbours a millisecond
pulsar (\S~\ref{section:1808_1998_outburst}), the three observed
X-ray bursts were scrutinized for potential burst oscillations
\cite{intzand2001}. A marginal detection of a 401 Hz oscillation was
made in the third burst. This result suggested that the burst
oscillations observed in the other, non-pulsating, neutron-star LMXBs
occur indeed at their neutron-star spin frequencies. This result has
been confirmed by the recent detection of burst oscillations during
the 2002 outburst of SAX J1808.4--3658
(\S~\ref{section:1808_2002_outburst}).

\subsection{The 1998 April outburst \label{section:1808_1998_outburst}}

On 1998 April 9, {\itshape RXTE}/PCA slew observations indicated that
SAX J1808.4--3658 was active again \cite{marshall1998}.  Using public
TOO observations of this source from 1998 April 11, it was discovered
\cite{wvdk1998} that coherent 401 Hz pulsations were present in the
persistent X-ray flux of the source, making it the first
accretion-driven millisecond X-ray pulsar discovered. After this
discovery, several more {\it RXTE}/PCA observations were made which
were used by several groups to study different aspects of the source.
I will only briefly mention those results and give references for
details.

A detailed analysis of the coherent timing behaviour showed
\cite{cm1998} that the neutron star was in a tight binary with a very
low-mass companion star in a $\sim$2-hr orbital period. Due to the
limited amount of data obtained during this outburst, only an upper
limit of $<7\times 10^{-13}$ Hz s$^{-1}$ could be obtained on the
pulse-frequency derivative \cite{cm1998}. Studies of the X-ray
spectrum \cite{gilfanov1998,hs1998,gierlinski2002} and the aperiodic
rapid X-ray variability \cite{wvdk1998_bbn} showed an object that,
apart from its pulsations, is remarkably similar to other LMXBs with
comparable luminosities (the atoll sources). There is apparent
modulation of the X-ray intensity at the orbital period, with a broad
minimum when the pulsar is behind the companion \cite{cm1998,hs1998}.
Cui et al.~\cite{cui1998} and Ford \cite{ford1999} reported on the
harmonic content, the energy dependency, and the soft phase lag of the
pulsations.

Another interesting aspect is that the source first showed a steady
decline in X-ray flux, which after $\sim$2 weeks suddenly accelerated
\cite{gilfanov1998,cui1998}. This behaviour has been attributed to the
fact that the source might have entered the 'propeller regime' in
which the accretion is centrifugally inhibited
\cite{gilfanov1998}. However, after the onset of the steep decline
the pulsations could still be detected \cite{cui1998} making this
interpretation doubtful. A week after the onset of this steep decline,
the X-ray flux levelled off \cite{cui1998,wang2001}, but as no further
{\itshape RXTE}/PCA observations were made, the X-ray behaviour of the
source at the end of the outburst remained unclear. The source might
have displayed a similar long-term episode of low-luminosity activity
as seen at the end of its 2000 and 2002 outbursts
(\S\S~\ref{section:1808_2000_outburst}-\ref{section:1808_2002_outburst}).

SAX J1808.4--3658 was not only detected and studied in X-rays but
also in the optical, the IR, and in the radio. The optical/IR
counterpart of SAX J1808.4--3658 (later named V4580 Sgr;
\cite{kazarovets2000}) was first discovered by Roche et al.
\cite{roche1998} and subsequently confirmed by Giles et al.
\cite{giles1998}. A detailed study of the optical behaviour during
this outburst was reported by Giles et al. \cite{giles1999} and Wang
et al. \cite{wang2001}. Both papers reported that the peak V
magnitude of the source was $\sim$16.7 and the source decayed in
brightness as the outburst progressed. The brightness of the source
levelled off at around V $\sim$ 18.5 (I $\sim$ 17.9) about $\sim$2
weeks after the peak of the outburst. It stayed at this level for at
least several  weeks before it further decreased in brightness. This
behaviour suggests that the source was indeed still active for a long
period after the main outburst.

It was also reported \cite{giles1999} that the optical flux was
modulated at the 2-hr orbital period of the system. Modelling the
X-ray and optical emission from the system using an X-ray-heated
accretion disk model, a $A_v$ of 0.68 and an inclination of $\cos i =
0.65$ were obtained \cite{wang2001}, resulting in a mass of the
companion star of 0.05--0.10 \mzon. Some of the IR observations were
too bright to be consistent with emission from the disk or the
companion star, even when considering X-ray heating. This IR excess
might be due to synchrotron processes, likely related to an outflow
or ejection of matter \cite{wang2001}. Such an event was also
confirmed by the discovery of the radio counterpart
\cite{gaensler1999}. The source was detected with a 4.8 GHz flux of
$\sim$0.8 mJy on 1998 April 27, but it was not detected at earlier or
later epochs.

\subsection{The 2000 January outburst \label{section:1808_2000_outburst}}

\begin{figure}
\includegraphics*[width=7.5cm]{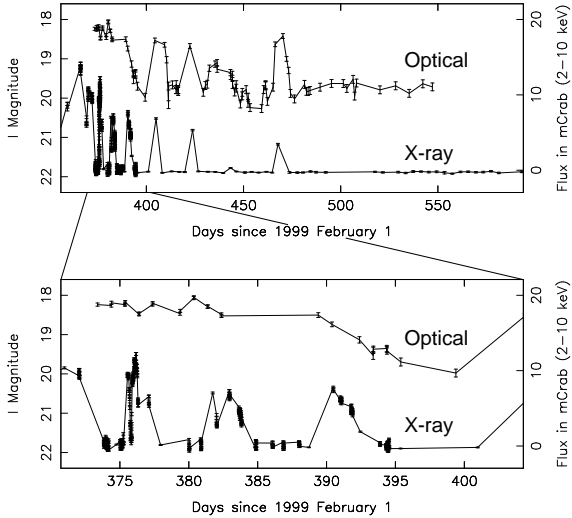}
\caption{
The {\it RXTE}/PCA \cite{wijnands2001} and the optical (I band) light
curves \cite{wachter2000} of \hbox{SAX J1808.4--3658} as observed
during its 2000 outburst. The optical data were kindly provided by
Stefanie Wachter.
\label{fig:1808_2000_optical}}
\end{figure}

On 2000 January 21, SAX J1808.4--3658 was again detected
\cite{wijnands2001} with the {\it RXTE}/PCA at a flux level of
$\sim$10--15 mCrab (2--10 keV), i.e. about a tenth of the peak fluxes
observed during the two previous outbursts. Using follow-up {\it
RXTE}/PCA observations, it was found that the source exhibited
low-level activity for several months (\cite{wijnands2001};
Fig.~\ref{fig:1808_2000_optical}). Due to solar constraints the source
could not be observed before January 21 but likely a true outburst
occurred before that date and only the end stages of this outburst was
observed. This is supported by the very similar behaviour of the
source observed near the end of its 2002 October outbursts
(\S~\ref{section:1808_2002_outburst}).

\begin{figure}
\includegraphics*[width=7.5cm]{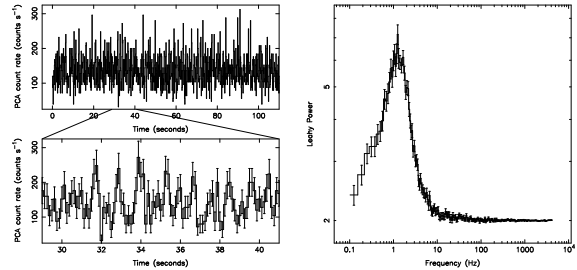}
\caption{
The violent 1 Hz flaring as observed during the 2000 outburst of SAX
J1808.4--3658. The left panels show the flaring in the light curve and
the right panel in the power spectrum.
\label{fig:1808_2000_1Hzflares}}
\end{figure}

During the 2000 outburst, SAX J1808.4--3658 was observed (using
{\itshape RXTE}) on some occasions at luminosities of $\sim10^{35}$
ergs s$^{-1}$, but on other occasions (a few days earlier or later) it
had luminosities of $\sim10^{32}$ ergs s$^{-1}$ (from {\itshape
BeppoSAX} and {\itshape XMM-Newton} observations
\cite{wijnands2002,wijnands2003}). This demonstrates that the source
exhibited extreme luminosity swings (a factor of $>1000$) on
timescales of days.  During the {\it RXTE} observations, it was also
found that on several occasions the source exhibited strong (up to
100\% rms amplitude) violent flaring behavior with a repetition
frequency of about 1 Hz (\cite{vdk2000_1808,wijnands2001_flaring};
Fig.\ref{fig:1808_2000_1Hzflares}). During this episode of low-level
activity, the pulsations at 401 Hz were also detected. The limited
amount of observing time and the low count rates of the source did
not allow for an independent determination of the binary orbital
parameters and the pulse-frequency derivative.

The source was again detected in optical, albeit at a lower brightness
than during the 1998 outburst \cite{wh2000}. This is consistent with
the lower X-ray activity seen for the source. The source was
frequently observed during this outburst and preliminary results were
presented by Wachter et al. \cite{wachter2000}. The main results are
presented in Figure~\ref{fig:1808_2000_optical} (reproduced with
permission from Stefanie Wachter). The optical and X-ray brightness of
the source are correlated at the end of the outburst, although one
optical flare (around day 435--440 in
Fig.~\ref{fig:1808_2000_optical}) was not accompanied by an X-ray
flare. However, the optical and X-ray observations were not
simultaneously, which means that a brief (of order a few days) X-ray
flare could have been missed. During the earlier stages of the
outburst, the X-ray and the optical behaviour of the source were not
correlated (Fig.~\ref{fig:1808_2000_optical} lower panel): the source
is highly variable in X-rays, but quite stable in optical with only
low amplitude variations. This stable period in the optical is very
similar to the episode of stable optical emission at the late stages
of the 1998 outburst, suggesting typical behaviour of the source.

\subsection{The 2002 October outburst \label{section:1808_2002_outburst}}

In 2002 October, the fourth outburst of SAX J1808.4--3658 was detected
\cite{markwardt2002_1808} and immediately a very extensive {\itshape
RXTE}/PCA observing campaign started. The main results are summarised
below.

\subsubsection{The X-ray light curve}

\begin{figure}
\includegraphics*[angle=-90,width=7.5cm]{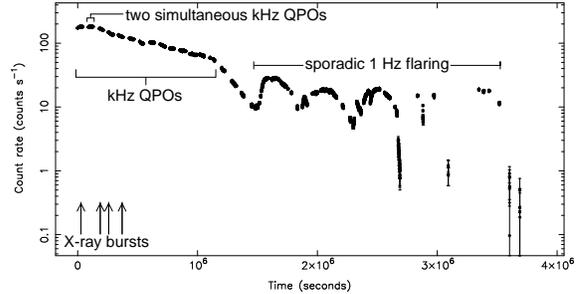}
\caption{
The light curve of SAX J1808.4--3658 during its 2002 outburst. The
count rate is per detector and for the energy range 2.5--18.1 keV. The
time is in seconds since the start of the first observation. It is
indicated in the figure when the different types of phenomena were
observed. \label{fig:1808_lc_2002}}
\end{figure}

The light curve for this outburst is shown in
Figure~\ref{fig:1808_lc_2002}. During the first few weeks, the source
decayed steadily, until the rate of decline suddenly increased, very
similar to what was observed during the 1998 outburst
(\S~\ref{section:1808_1998_outburst}). About five days later the X-ray
count rate rapidly increased again until it reached a peak of about a
tenth of the outburst maximum. After that the source entered a state
in which the count rate rapidly fluctuated on time scales of days to
hours, very similar to the 2000 low-level activity
(\S~\ref{section:1808_2000_outburst}). This outburst light curve is
the most detailed one seen for this source and it exhibits all
features seen during the previous three outbursts of the source (the
initial decline, the increase in the decline rate, the long-term
low-level activity), demonstrating that this is typical source
behaviour.

\subsubsection{The pulsations}

The pulsations could be detected at all flux levels with an amplitude
of 3\%--10\%.  The pulsar was spinning down at a constant rate (mean
spin-down rate of $2\times 10^{-13}$ Hz s$^{-1}$;
\cite{chakrabarty2003}), despite a large dynamic range of X-ray flux.
The magnitude of the pulse-frequency derivative exceeds the maximum
value expected from accretion torques by a factor of 5. The timing
history also contains a small glitch with a very rapid recovery time
scale.  There was no evidence for a 200.5 Hz subharmonic in the data
(upper limit of 0.38\% of the signal at 401
Hz;\cite{wijnands2003_nature}) confirming the interpretation of 401 Hz
as the pulsar spin frequency. A more detailed analysis will be
presented elsewhere \cite{morgan2003}.

\subsubsection{X-ray bursts and burst oscillations}

During the first five days of the outburst, four type-I X-ray bursts
were detected. During the rise and decay  of each burst, but not
during the peak, burst oscillations were observed
\cite{chakrabarty2003}: the frequency in the burst tails was constant
and identical to the spin frequency, while the oscillation in the
burst rise showed evidence for a very rapid frequency drift of up to 5
Hz. This frequency behaviour and the absence of oscillations at the
peak of the bursts is similar to the burst oscillations seen in other,
non-pulsating neutron star LMXBs, demonstrating that indeed the
burst-oscillations occur at the neutron-star spin frequency in all
sources. As a consequence, the spin frequency is now known for 16
LMXBs (11 burst-oscillations sources and 5 pulsars) and the highest
spin frequency is 619 Hz. The sample of burst-oscillation sources was
used to demonstrated that neutron stars in LMXBs spin well below the
break-up frequency for neutron stars. This could suggest that the
neutron stars are limited in their spin frequencies, possible due to
the emission of gravitational radiation \cite{chakrabarty2003}.

\subsubsection{The kHz QPOs}

\begin{figure}[t]
\includegraphics[width=7.5cm]{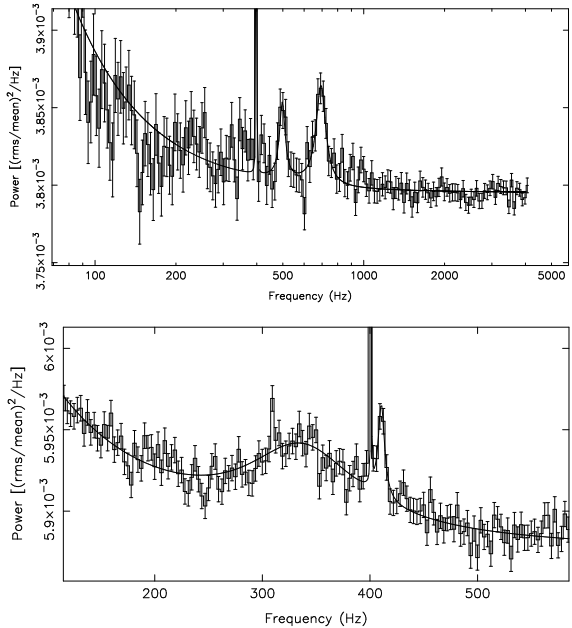}
\caption{The top panel shows the two simultaneous kHz QPOs discovered for SAX
J1808.4--3658 during its 2002 outburst. The bottom panel shows the
enigmatic 410 Hz QPO also seen for this source during this outburst.
The figures are adaptations from the figures in
\cite{wijnands2003_nature} \label{fig:1808_2002_kHzQPOs}}
\end{figure}

Wijnands et al. \cite{wijnands2003_nature} reported on the discovery
of two simultaneous kHz QPOs during the peak of the outburst, with
frequencies of $\sim$700 and $\sim$500 Hz (during the meeting, this
result was presented by Michiel van der Klis;
Fig.~\ref{fig:1808_2002_kHzQPOs} top). This was the first detection of
twin kHz QPOs in a source with a known spin-frequency.  The frequency
separation of those two kHz QPOs is only $\sim$200 Hz, significantly
below the 401 Hz expected in the beat-frequency models proposed to
explain the kHz QPOs. Therefore, those models are falsified by the
discovery of kHz QPOs in SAX J1808.4--3658. The fact that the peak
separation is approximately half the spin frequency suggests that the
kHz QPOs are indeed connected to the neutron-star spin frequency,
albeit in a way not predicted by any model. The lower-frequency kHz
QPO was only seen during the peak of the outburst (2002 October 16)
but the higher-frequency kHz QPO could be traced throughout the main
part of the outburst \cite{wijnands2003_nature}.  Besides the twin kHz
QPOs, a third kHz QPO was found with frequencies ($\sim$410~Hz) just
exceeding the pulse frequency \cite[Fig.~\ref{fig:1808_2002_kHzQPOs}
bottom]{wijnands2003_nature}. The nature of this QPO is unclear but it
might be related to the side-band kHz QPO seen in several other
sources \cite{jonker2000}.

Wijnands et al. \cite{wijnands2003_nature} pointed out that there seem
to be two classes of neutron-star LMXBS: the 'fast' and the 'slow'
rotators. The fast rotators have spin frequencies $\ga$400 Hz and the
frequency separation between the kHz QPOs is roughly equal to half the
spin frequency. In contrast, the slow rotators have spin frequencies
below $\la$400 Hz and a frequency separation roughly equal to the spin
frequency. These new kHz QPO results have already spurred new
theoretical investigations in the kHz QPO nature, involving spin
induced resonance in the disk
\cite{wijnands2003_nature,kluzniak2003,lm2003}.

\subsubsection{The low-frequency QPOs}

\begin{figure}
\includegraphics*[width=7.5cm]{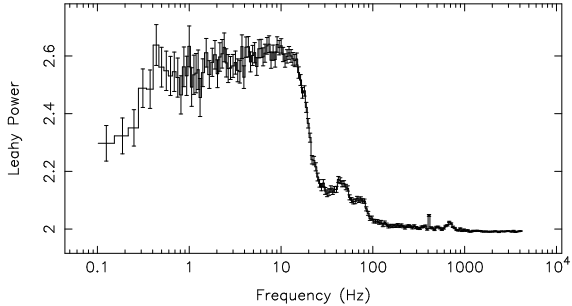}
\caption{
The broad-band noise and the 40--80 Hz QPOs seen during the 2002
outburst of SAX J1808.4--3658. The upper kHz QPO and the pulsations
can also been seen. \label{fig:1808_2002_noise}}
\end{figure}

During the peak of the outburst and in its subsequent decay,
broad-noise and QPOs with frequencies between 10 and 80 Hz were
detected in the power spectra (Fig.~\ref{fig:1808_2002_noise}).
Similar phenomena have been observed in other non-pulsating systems
and are likely related to the noise components seen in SAX
J1808.4--3658. For a discussion about the low-frequency QPOs and their
connection to the kHz QPOs, I refer to the contribution by Steve van
Straaten in this proceedings.

\subsubsection{The violent 1 Hz flaring}

Violent flaring was observed on many occasions at a $\sim$1 Hz
repetition frequency during the late stages of the 2002 outburst
(Fig.~\ref{fig:1808_lc_2002}), similar to what had been observed
during the 2000 outburst. This proves that also this violent flaring
is a recurrent phenomenon and can likely be observed every time the
source is in this prolonged low-level activity state. The mechanism
behind these violent flares is not yet known and a detailed analysis
of this phenomenon is in progress.

\subsubsection{Observations at other wavelengths}

Rupen et al. \cite{rupen2002} reported the detection of the source at
radio wavelengths. On 16 October 2002, they found a 0.44-mJy source at
8.5 GHz and a day later, the source was detected at 0.3 mJy. Monard
\cite{monard2002} reported that on 16 October 2002 the optical
counterpart was detectable again at magnitudes similar to those
observed at the peak of the 1998 outburst.

\subsection{SAX J1808.4--3658 in quiescence}

In quiescence, SAX J1808.4--3658 has been observed on several
occasions with the {\it BeppoSAX} and {\it ASCA} satellites
\cite{stellaetal2000,daw2000,wijnandsetal2002_bepposax}. The
source was very dim in quiescence, with a luminosity close to or lower
than $10^{32}$ \Lunit.  Due to the low number of source photons
detected, these luminosities had large errors and no information could
be obtained on the spectral shape or possible variability in
quiescence.  Due to the limited angular resolution of {\it BeppoSAX},
doubts were raised as to whether the source detected by this satellite
was truly SAX J1808.4--3658 or an unrelated field source
\cite{wijnandsetal2002_bepposax}. Campana et
al. \cite{campanaetal2002} reported on a quiescent observation of the
source performed with {\it XMM-Newton} which resolved this issue. They
detected the source at a luminosity of $5\times10^{31}$ \Lunit~and
found that the field around SAX J1808.4--3658 is rather crowded with
weak sources. Two such sources are relatively close to SAX
J1808.4--3658 and might have conceivably caused a systematic
positional offset during the {\it BeppoSAX} observations of SAX
J1808.4--3658. Very likely the source was indeed detected during those
observations.

Using {\itshape XMM-Newton}, Campana et al. \cite{campanaetal2002}
obtained enough photons to extract a quiescent X-ray spectrum, which
was not dominated by the same thermal component seen in other
quiescent neutron star transients; such a thermal component is thought
to be due to the cooling of the neutron star in-between
outbursts. However, the spectrum of SAX J1808.4--3658 was dominated by
a power-law shaped component.  The non-detection of the thermal
component was used to argue that the neutron star was anomalously
cool, possibly due to enhanced core cooling processes
\cite{campanaetal2002}.  It has been argued
\cite{stellaetal2000,campanaetal2002} that the propeller mechanism,
which might explain (some of) the hard X-ray emission in quiescence,
is likely not active since this mechanism is expected to stop
operating at luminosities $<10^{33}$ \Lunit, because at those
luminosities the source should turn on as a radio pulsar. Instead, it
was proposed that the quiescent X-rays originate in the shock between
the wind of a turned-on radio pulsar and the matter out-flowing from
the companion \cite{stellaetal2000,campanaetal2002}.  The quiescent
X-rays could also be due to direct dipole radiation from the radio
pulsar \cite{dsb2003}.

The quiescent optical counterpart of SAX J1808.4--3658 was studied by
Homer et al. \cite{homer2001}. They reported that on 1999 August 10
the orbital modulation was still present in white light observations
(estimated V magnitude of $\sim$20), with an semi-amplitude of
$\sim$6\%. It has the same phasing and approximately sinusoidal
modulation as seen during outburst, and with photometric minimum when
the pulsar is behind the companion star.  During observations taken in
July 2000 the quiescent counterpart was even fainter and no
significant orbital modulation could be detected. Using these results,
it has been suggested that the optical properties of SAX J1808.4--3658
in quiescence are evidence of an active radio pulsar
\cite{burderi2003}.

\section{XTE J1751--305}

The second accretion-driven millisecond pulsar (XTE J1751--305) was
discovered on 2002 April 3 \cite{markwardt2002}. Its spin frequency is
435 Hz and the neutron star is in a very small binary with an orbital
period of only 42 minutes. The timing analysis of the pulsations gave
a minimum mass for the companion star of 0.013
\mzon~and a pulse-frequency derivative of $<3 \times 10^{-13}$ Hz
s$^{-1}$. Assuming that the mass transfer in this binary system was
driven by gravitational radiation, the distance toward the source
could be constrained to at least 7 kpc and the orbital inclination to
30\degr--85\degr, resulting in a companion mass of 0.013--0.035 \mzon,
suggesting a heated helium dwarf \cite{markwardt2002}.

The source reached a peak luminosity of $>$2$\times 10^{37}$ \Lunit,
an order of magnitude brighter than the peak luminosity of SAX
J1808.4--3658. However, the outburst was very short with an e-folding
time of only $\sim$7 days (compared to $\sim14$ days for SAX
J1808.4--3658) resulting in a low outburst fluence of only $\sim2.5
\times 10^{-3}$ ergs cm$^{-2}$ \cite{markwardt2002}. A potential
re-flare was seen two weeks after the end of the outburst during which
also a type-I X-ray burst was seen. Preliminary analysis of the burst
indicated that the burst did not come from XTE J1751--305 but from
another source in the field-of-view. This was later confirmed
\cite{intzandterzan} and the burst likely originated from the bright
X-ray transient in Terzan 6 (however, this transient did not produce
the re-flare, which can still have come from XTE
J1751--305). {\itshape Chandra} also briefly observed the source,
resulting in an arcsecond position, and a previous outburst in 1998
June was detected using archival {\it RXTE}/ASM data
\cite{markwardt2002}, suggesting a tentative recurrence time of
$\sim$3.8 years.

Miller et al. \cite{miller2002} reported on high spectral resolution
data of the source obtained with {\itshape XMM-Newton} to search for
line features in the X-ray spectrum. However, they only detected a
continuum spectrum dominated by a hard power-law shaped component
(power-law index of $\sim1.44$) but with a 17 \% contribution to the
0.5--10 keV flux by a soft thermal (black-body) component with
temperature of $\sim1$ keV. Searches for the optical and near-infrared
counterparts were performed but no counterparts were found
\cite{jonker2003}, likely due to the high reddening toward the
source. These non-detections did not constrain any models for the
accretion disk or possible donor stars.

\section{XTE J0929--314}

The third accretion-driven millisecond X-ray pulsar XTE J0929--314
was already detected with the {\itshape RXTE}/ASM on 13 April 2002
\cite{remillard2002} but was only found to be harbouring a
millisecond pulsar with a pulsations frequency of 185 Hz on 2 May
when observations of the source were made using the {\it RXTE}/PCA
\cite{remillardetal2002}. Galloway et al. \cite{galloway2002}
reported on the detection of the 44-min orbital period of the system
which is remarkably similar to that of XTE J1751--305. A minimum mass
of 0.008 \mzon was obtained for the companion star and a
pulse-frequency derivative of ($-9.2\pm0.4)\times 10^{-14}$ Hz
s$^{-1}$. Galloway et al. \cite{galloway2002} suggested that this spin
down torque may arise from magnetic coupling to the accretion disk, a
magneto-hydrodynamic wind, or gravitational radiation from the rapidly
spinning neutron star.  Assuming gravitational radiation as the
driving force behind the mass transfer, Galloway et al.
\cite{galloway2002} found a lower limit to the distance of 6 kpc. 
Juett et al. \cite{juett2003} obtained high resolution spectral data
using the {\itshape Chandra} gratings. Again the spectrum is well
fitted by a power-law plus a black body component, with a power-law
index of 1.55 and a temperature of 0.65 keV. Similar to XTE
J1751--305, no emission or absorption features were found. No orbital
modulation of the X-ray flux was found implying an upper limit on the
inclination of 85\degr.

Greenhill et al. \cite{greenhill2002} reported the discovery of the
optical counterpart of the system with a V magnitude of 18.8 on 1 May
2002. Castro-Tirado et al. \cite{ct2002} obtained optical spectra of
the source on May 6--8 in the range 350--800 nm and found emission
lines from the C III - N III blend and H-alpha, which were superposed
on a blue continuum. These optical properties are typical of X-ray
transients during outburst.  Rupen et al. \cite{rupen2002_0929}
discovered the radio counterpart of the source using the VLA with 4.86
GHz flux of 0.3--0.4 mJy.

\section{XTE J1807--294}

The fourth millisecond X-ray pulsar XTE J1807--294 with a frequency of
191 Hz, was discovered on 21 February 2003
\cite{markwardt2003_1807}. The peak flux was only 58 mCrab (2--10 keV,
measured on 21 February). The orbital period was determined
\cite{markwardt2003_atel} to be $\sim40$ minutes making it the
shortest period of all accretion-driven millisecond pulsars now
known. Using a {\itshape Chandra} observation, Markwardt et
al. \cite{markwardt2003_atel} reported the best known position of the
source. Using the {\it RXTE}/PCA data, kHz QPOs have been detected for
this system and the results obtained from a full analysis of those
data will be reported elsewhere \cite{markwardt2003_prep}.  Campana et
al. \cite{campana2003_1807} reported on a {\itshape XMM-Newton}
observation of this source taken on 22 March 2003. Assuming a distance
of 8 kpc, the 0.5--10 keV luminosity during that observation was
$2\times10^{36}$ \Lunit. They could detect the pulsations during this
observation with a pulsed fraction of 5.8\% in the 0.3--10 keV band
(increasing with energy) and a nearly sinusoidal pulse profile (see
also \cite{kk2003}).  The spectral data are well fit by a continuum
model, assumed to be an absorbed Comptonisation model plus a soft
component. The latter component only contributed 13\% to the
flux. Again no emission or absorption lines were found.  No detections
of the counterparts of the system at other wavelengths have so far
been reported.

\section{XTE J1814--338}

\begin{figure}
\begin{center}
\includegraphics*[width=6cm]{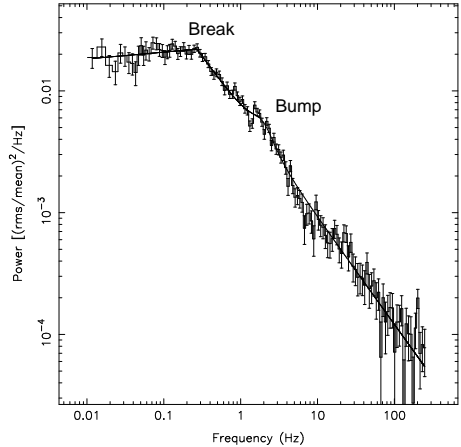}
\caption{
The broad-band noise observed for \hbox{XTE J1814--338}
\cite{wh2003}. \label{fig:1814_pds}}
\end{center}
\end{figure}

The fifth system (XTE J1814--338) was discovered on 5 June 2003 and
has a pulse frequency of 314 Hz \cite{markwardt2003_1814}, with an
orbital period of 4.3 hr and a minimum companion mass of 0.15 \mzon
\cite{markwardt2003_1814_atel}. This 4.3 hr orbital period makes it
the widest binary system among the accretion-driven millisecond
pulsars and also the one most similar to the general population the
low-luminosity neutron star LMXBs (the atoll sources).  Many type-I
X-ray bursts were seen during which burst oscillations were found with
a frequency consistent with the neutron star spin frequency
\cite{markwardt2003_1814_atel,strohmayer2003}. A distance of $\sim$8
kpc was obtained from the only burst which likely reached the
Eddington luminosity. The burst oscillations are strongly frequency
and phase locked to the persistent pulsations (as was also seen for
SAX J1808.4--3658; \cite{chakrabarty2003}) and two bursts exhibited
evidence for a frequency decrease of a few tenths of a Hz during the
onset of the burst, suggesting a spin down. Strohmayer et
al. \cite{strohmayer2003} also reported on the detection of the first
harmonic of the burst oscillations, which is the first time that this
has been found for any burst-oscillation source. This harmonic could
arise from two hot-spots on the surface, but they suggested that if
the burst oscillations arise from a single bright region, the strength
of the harmonic would suggest that the burst emission is beamed
(possible due to a stronger magnetic field strength than in
non-pulsating LMXBs).

Wijnands \& Homan \cite{wh2003} analysed the {\itshape RXTE}/PCA data
of the source obtained between 8 and 11 June 2003. The overall shape
of the 3-60 keV power spectrum is dominated by a strong broad
band-limited noise component (Fig.~\ref{fig:1814_pds}), which could be
fitted by a broken power-law model with a broad bump superimposed on
it at frequencies above the break frequency. These characteristics
make the power spectrum of XTE J1814-338 very similar to that observed
in the non-pulsing low-luminosity neutron-star LMXBs (the atoll
sources) when they are observed at relatively low X-ray luminosities
(i.e., in the so-called island state). This is consistent with the
hard power-law X-ray spectrum of the source reported by Markwardt et
al. \cite{markwardt2003_1814_atel}. No kHz QPOs were found, although
the upper limits were not very stringent.

Wijnands \& Reynolds \cite{wr2003} reported that the position of XTE
J1814--338 was consistent with the {\itshape EXOSAT} slew source EXMS
B1810--337 which was detected on 2 September 1984. If XTE J1814-338
can indeed be identified with EXMS B1810-337, then its recurrence time
can be inferred to be less than 19 years but more than 4.5 years (the
time since the {\itshape RXTE}/PCA bulge scan observations started in
February 1999), unless the recurrence time of the source varies
significantly.

Krauss et al. \cite{krauss2003} reported the best position of the
source as obtained using {\itshape Chandra} and on the detection of
the likely optical counterpart of the source (with magnitudes of B =
17.3 and R = 18.8 on June 6). Steeghs \cite{steeghs} reported on
optical spectroscopy of this possible counterpart and prominent
hydrogen and helium emission lines were detected, confirming the
connection between the optical source and XTE J1814--338.

\section{Theoretical work}

The lack of space for this review does not allow me to go into detail
on the theoretical papers published on accretion-driven millisecond
pulsars. Here, I will only briefly list some of those papers, which
mostly focus on SAX J1808.4--3658 since the other four systems have
only been found very recently. Since the discovery of SAX
J1808.4--3658, several studied have tried to constrain the properties
(i.e., radius, mass, magnetic field strength) of the neutron star in
this system \cite{bk1998,pc1999,sudipb2001}, while others proposed
that the compact object is not a neutron star at all, but instead a
strange star (see, e.g., \cite{li,datta,zdunik2000} and the references
in those papers). Other studies focused on the evolutionary history of
this system \cite{ea1999} or on the nature of the companion star
\cite[who suggested a brown dwarf companion star]{bc2001}.

The discovery of the accretion-driven millisecond pulsars raises the
important question as to why those systems are different from the
other neutron star LMXBs for which not pulsations have been found.
Cumming et al. (\cite{cumming2001}; see also \cite{rck2002}) suggested
that the low time-averaged accretion rate of SAX J1808.4--3658 might
explain why this source is a pulsar. Although the remaining four
pulsars were not know at the time of writing of that paper, the same
arguments can also be used for those systems: when the time-averaged
accretion rate is sufficiently high, the neutron star magnetic field
might be buried by the accreted matter and does not have time to
dissipate through the accreted material. However, for the pulsars the
time-averaged accretion rate is sufficiently low that this can indeed
happen and therefore those systems have a magnetic field which is
still strong enough to disturb the flow of the accreted
matter. However, more neutron-star LMXBs with low time-averaged
accretion rate have to be found and studied in detail to investigate
whether they all harbour a millisecond pulsar or if there are also
systems with low time-averaged accretion rates that do not harbour a
millisecond pulsar. In the latter scenario, the screening idea might
only be part of the explanation and alternative ideas need to be
explored (see, e.g., \cite{titarchuk2002}).

\vspace{0.25cm}
{\itshape Acknowledgements.} I thank Stefanie Wachter for kindly
providing the optical data used in Figure~\ref{fig:1808_2000_optical}
and Jeroen Homan for carefully reading an earlier version of this
manuscript.


\begin{thebibliography}{}

\bibitem{bvdh1991}Bhattacharya, D. \& van den Heuvel, E.P.J. 
1991, Ph.R. 203, 1

\bibitem{v1994}Vaughan, B.A. et al. 1994, ApJ, 435, 362

\bibitem{strohmayer1996}Strohmayer, T.E. et al. 1996, ApJ, 469, L9

\bibitem{vdk1996}Van der Klis, M. et al. 1996, ApJ, 469, L1

\bibitem{vdk2000}Van der Klis, M. 2000, ARA\&A, 38, 717

\bibitem{sb2003}Strohmayer, T. \& Bildsten, L. 2003, To 
appear in 'Compact Stellar X-ray sources', eds. W.H.G. Lewin \&
M. van der Klis, Cambridge University Press (astro-ph/0301544)

\bibitem{wvdk1998}Wijnands, R. \& van der Klis, M. 1998, Nature, 
394, 344

\bibitem{markwardt2002}Markwardt, C.B. et al. 2002, ApJ, 575, L21

\bibitem{galloway2002}Galloway, D.K. et al. 2002, ApJ, 576, L137

\bibitem{markwardt2003_1807}Markwardt, C.B. et al. 2003, IAUC 8080

\bibitem{markwardt2003_1814}Markwardt, C.B. et al. 2003, IAUC 8144

\bibitem{intzand1998}In 't Zand, J.J.M. et al. 1998, A\&A, 331, L25

\bibitem{intzand2001}In 't Zand, J.J.M. et al. 2001, A\&A, 372, 916

\bibitem{heise1999}Heise, J. et al. 1999, ApL\&C 38, 297

\bibitem{revnivtsev2003}Revnivtsev, M.G. 2003, AstL, 29, 383

\bibitem{marshall1998}Marshall, F.E. 1998, IAUC 6876

\bibitem{cm1998}Chakrabarty, D. \& Morgan, E. H. 1998, Nature, 394, 346

\bibitem{gilfanov1998}Gilfanov, M. et al. 1998, A\&A, 338, L83

\bibitem{hs1998}Heindl, W.A. \& Smith, D.M. 1998, ApJ, 506, L35

\bibitem{gierlinski2002}Gierlinski, M. et al. 2002, MNRAS, 331, 141

\bibitem{wvdk1998_bbn}Wijnands, R. \& van der Klis, M. 1998, ApJ, 507, L63

\bibitem{cui1998}Cui, W. et al. 1998, ApJ, 504, L27

\bibitem{ford1999}Ford, E.C. 1999, ApJ, 519, L73

\bibitem{wang2001}Wang, Z. et al. 2001, ApJ, 563, L61

\bibitem{kazarovets2000}Kazarovets, E.V. et al. 2000, IBVS, 4870

\bibitem{roche1998}Roche, P. et al. 1998, IAUC 6885

\bibitem{giles1998}Giles, A.B. et al. 1998, IAUC 6886

\bibitem{giles1999}Giles, A.B. et al. 1999, MNRAS, 304, 47

\bibitem{gaensler1999}Gaensler, B.M. et al. 1999, ApJ, 522, L117

\bibitem{wijnands2001}Wijnands, R. et al. 2001, ApJ, 560, 892

\bibitem{wijnands2002}Wijnands, R. et al. 2002, ApJ, 571, 429

\bibitem{wijnands2003}Wijnands, R. 2003, ApJ, 588, 425

\bibitem{vdk2000_1808}Van der Klis, M. et al. 2000, IAUC, 7358

\bibitem{wijnands2001_flaring}Wijnands, R. 2001, Adv. Space Res. 28, 469

\bibitem{wh2000}Wachter, S. \& Hoard, D.W. 2000, IAUC 7363

\bibitem{wachter2000}Wachter, S. et al. 2000, HEAD 32, 24.15

\bibitem{markwardt2002_1808}Markwardt, C.B. et al. 2002, IAUC 7993

\bibitem{chakrabarty2003}Chakrabarty, D. et al. 2003, Nature, 424, 42

\bibitem{wijnands2003_nature}Wijnands, R. et al.2003, Nature, 424, 44

\bibitem{morgan2003}Morgan, E.H. et al. 2003, ApJ in preparation

\bibitem{jonker2000}Jonker, P.G. et al. 2000, ApJ, 540, L29

\bibitem{kluzniak2003}Klu\'zniak, W. et al. 2003, astro-ph/0308035

\bibitem{lm2003}Lamb, F.K. \& Miller, M.C. 2003, ApJ Letters, 
submitted (astro-ph/0308179)


\bibitem{rupen2002}Rupen, M. et al. 2002, IAUC 7997

\bibitem{monard2002}Monard, B. 2002, VSNet alert 7550


\bibitem{stellaetal2000}Stella, L. et al. 2000, ApJ, 537, L115

\bibitem{daw2000}Dotani, T. et al. 2000, 
ApJ, 543, L145

\bibitem{wijnandsetal2002_bepposax}Wijnands, 
R. et al. 2002, ApJ, 571, 429

\bibitem{campanaetal2002}Campana, S. et al. 2002, ApJ, 575, L15)

\bibitem{dsb2003}Di Salvo, T. \& Burderi, L. 2003, A\&A, 397, 723

\bibitem{homer2001}Homer, L. et al. 2001, MNRAS, 325, 1471

\bibitem{burderi2003}Burderi, L. et al. 2003, A\&A, 404, 43

\bibitem{intzandterzan}In 't Zand, J.J.M. et al. 2003, A\&A
submitted (astro-ph/0307044)

\bibitem{miller2002}Miller, J.M. et al. 2003, ApJ, 583, L99

\bibitem{jonker2003}Jonker, P.G. et al. 2003, MNRAS, 344, 201

\bibitem{remillard2002}Remillard, R.A. 2002, IAUC 7888

\bibitem{remillardetal2002}Remillard, R.A. et al. 2002, IAUC 7893

\bibitem{juett2003}Juett, A.M. et al. 2003, ApJ, 587, 754

\bibitem{greenhill2002}Greenhill, J.G. et al. 2002, IAUC 7889

\bibitem{ct2002}Castro-Tirado, A.J. et al. 2002, IAUC 7895

\bibitem{rupen2002_0929}Rupen, M.P. et al. 2002, IAUC 7893

\bibitem{markwardt2003_atel}Markwardt, C.B. et al. 2003, ATEL 127

\bibitem{markwardt2003_prep}Markwardt, C.B. et al. 2003, ApJ in preparation

\bibitem{campana2003_1807}Campana, S. et al. 2003, ApJ Letters in
 press (astro-ph/0307174)

\bibitem{kk2003}Kirsch, M.G.F. \& Kendziorra, E. 2003, ATEL 148

\bibitem{markwardt2003_1814_atel}Markwardt, C.B. et al. 2003, ATEL 164

\bibitem{strohmayer2003}Strohmayer, T.E. et al. 2003, ApJ Letters in
press (astro-ph/0308353)

\bibitem{wh2003}Wijnands, R. \& Homan, J. 2003, ATEL  165

\bibitem{wr2003}Wijnands, R. \& Reynolds, A. 2003, ATEL 166

\bibitem{krauss2003}Krauss, M.I. et al. 2003, IAUC 8154

\bibitem{steeghs}Steeghs, D. 2003, IAUC 8155

\bibitem{bk1998}Burderi, L. \& King. A.R. 1998, ApJ, 505, L135

\bibitem{pc1999}Psaltis, D. \& Chakrabarty, D. 1999, ApJ, 521, 332

\bibitem{sudipb2001}Bhattacharyya, S. 2001, ApJ, 554, 185

\bibitem{li}Li, X.-D. et al. 1999, Ph. Rev. Letters 83, 3776

\bibitem{datta}Datta, B. et al. 2000, A\&A, 355, 19

\bibitem{zdunik2000}Zdunik, J.L. et al. 2000, A\&A, 359, 143

\bibitem{ea1999}Ergma, E. \& Antipova, J. 1999, A\&A, 343, L45

\bibitem{bc2001}Bildsten, L. \& Chakrabarty, D. 2001, ApJ, 557, 292

\bibitem{cumming2001}Cumming, A. et al. 2001, ApJ 557, 958

\bibitem{rck2002}Rai Choudhuri, A. \& Konar, S. 2002, MNRAS, 332, 933

\bibitem{titarchuk2002}Titarchuk, L. et al. 2002, ApJ, 576, L49

\end{thebibliography}
\end{document}